\documentclass[
    ,final            
  ]
  {aipproc}

\layoutstyle{8x11single}

\begin{document}

\title{Swift follow-up of the gigantic TeV outburst\\of PKS $2155-304$ in 2006}


\author{L. Foschini}{
  address={INAF/IASF-Bologna, Via Gobetti 101, 40129, Bologna (Italy)}
}

\author{G. Ghisellini}{
  address={INAF/Osservatorio Astronomico di Brera, Via Brera 28, 20121 Milano (Italy)}
}

\author{F. Tavecchio}{
  address={INAF/Osservatorio Astronomico di Brera, Via Brera 28, 20121 Milano (Italy)}
}

\author{A. Treves}{
  address={Dipartimento di Scienze, Universit\`a dell'Insubria, Via Vallegio 11, 22100 Como (Italy)}
}

\author{L. Maraschi}{
  address={INAF/Osservatorio Astronomico di Brera, Via Brera 28, 20121 Milano (Italy)}
}

\author{M. Gliozzi}{
  address={George Mason University, Department of Physics and Astronomy, School of Computational Sciences, Mail Stop 3F3, 4400 University Drive, Fairfax, VA 22030, USA}
}

\author{C.M. Raiteri}{
  address={INAF/Osservatorio Astronomico di Torino, Via Osservatorio 20, 10025 Pino Torinese (Italy)}
}

\author{M. Villata}{
  address={INAF/Osservatorio Astronomico di Torino, Via Osservatorio 20, 10025 Pino Torinese (Italy)}
}

\author{E. Pian}{
  address={INAF/Osservatorio Astronomico di Trieste, Via G.B. Tiepolo 11, 34131 Trieste (Italy)}
}

\author{G. Tagliaferri}{
  address={INAF/Osservatorio Astronomico di Brera, Via Brera 28, 20121 Milano (Italy)}
}

\author{G. Tosti}{
  address={Osservatorio Astronomico, Universit\`a di Perugia, Via B. Bonfigli, 06126 Perugia (Italy)}
}

\author{R.M. Sambruna}{
  address={NASA/Goddard Space Flight Center, Code 661, Greenbelt, MD 20771, USA}
}

\author{G. Malaguti}{
  address={INAF/IASF-Bologna, Via Gobetti 101, 40129, Bologna (Italy)}
}

\author{G. Di Cocco}{
  address={INAF/IASF-Bologna, Via Gobetti 101, 40129, Bologna (Italy)}
}

\author{P. Giommi}{
  address={ASI Science Data Centre, Via G. Galilei, 00044 Frascati, (Italy)}
}

\begin{abstract}
At the end of July 2006, the blazar PKS $2155-304$ ($z=0.116$) underwent a strong outburst observed at TeV energies by HESS (up to $17$ Crab flux level at $E>200$~GeV). The \emph{Swift} satellite followed the evolution of the source for about one month. The data analysis -- reported in another paper -- has shown that, despite the violent activity at TeV energies, the synchrotron energy distribution increased in normalization, but only with a small shift in frequency. In the present work, we fit the broad-band spectrum with a log-parabolic model, to search for indications of intrinsic curvature, which in turn is usually interpreted as a signature of energy-dependent acceleration mechanisms of electrons.
\end{abstract}

\maketitle

\section{Introduction}
The HESS Cerenkov Telescope alerted the astronomical community on July 27, 2006 about an increase of activity of the blazar PKS $2155-304$ (z=0.116) in the TeV energy range \cite{hess:2006}. The same night the blazar displayed flares up to 17 Crab (E > 200 GeV) on time scales of 5 minutes \cite{hess:2007}. An immediate X-ray/UV/Optical follow-up with the \emph{Swift} satellite was activated and the analysis of data is reported in \cite{Foschini:2007}. A prompt increase of the X-ray flux without large spectral changes was revealed. Interestingly, the frequency of the synchrotron peak remained at values similar to those observed in the past, with low TeV activity \cite{Chiappetti:1999}. Further details can be found in \cite{Foschini:2007}. However, the event was so spectacular and anomalous that it deserves further attention.

\section{Spectral analysis with the log-parabolic model}
After an initial spectral fit of the X-ray data with power-law (PL) and broken power-law (BPL) models, we tried a log-parabolic model, which was already successfully applied to other blazars (e.g. \cite{Giommi:2002}, \cite{Massaro:2006}, \cite{Tramacere:2006}). This model is represented by the relationship:

\begin{equation}
	F(E) = K E^{-a-b\log(E)}
\end{equation}

\noindent where $E$ is the energy, $K$ the normalization, $a$ is the spectral slope (photon index) at 1 keV, and $b$ is the curvature. We applied this model to the \emph{Swift} observations with more than $100$ s of exposure on the X-ray Telescope (XRT), which are listed in Table 1 of \cite{Foschini:2007}. The results of the fits are reported in Table 1, while the values for $a$ and $b$ versus the normalization $K$ are shown in Fig. 1.
The fits with the log-parabolic model are generally comparable to the values obtained by using single or broken power law models reported in \cite{Foschini:2007}. In a few cases the log-parabolic model gives a better fit than the PL and BPL. In other cases, the resulting curvature is consistent with 0, i.e. the log-parabola mimics a PL. However, the limited photon statistic does not allow us to draw firm conclusions.

\begin{table}
\begin{tabular}{lcccc}
\hline
  \tablehead{1}{c}{b}{Date}
  & \tablehead{1}{c}{b}{a}
  & \tablehead{1}{c}{b}{b}
  & \tablehead{1}{c}{b}{F\tablenote{Observed flux in the $0.3-10$ keV energy band [$10^{-10}$ erg/cm$^2$/s].}}   
  & \tablehead{1}{c}{b}{$\tilde{\chi}^2$/dof}   \\
\hline
Apr 16    & $2.4\pm 0.1$   & $0.4\pm 0.3$   & $0.85$ & $1.28/40$ \\
Apr 26    & $2.4\pm 0.1$   & $<0.61$        & $1.27$ & $1.03/19$ \\
Jul 29/31 & $2.48\pm 0.01$ & $0.44\pm 0.04$ & $3.40$ & $1.16/320$ \\
Aug 1     & $2.63\pm 0.06$ & $0.48\pm 0.18$ & $2.70$ & $0.95/92$ \\
Aug 2     & $2.61\pm 0.02$ & $0.42\pm 0.07$ & $2.44$ & $1.00/196$ \\
Aug 3     & $2.46\pm 0.02$ & $0.45\pm 0.07$ & $2.83$ & $1.32/206$ \\
Aug 5     & $2.69\pm 0.06$ & $0.43\pm 0.19$ & $1.89$ & $0.80/92$ \\
Aug 6     & $2.64\pm 0.09$ & $<0.29$        & $1.68$ & $0.93/51$ \\
Aug 8     & $2.64\pm 0.06$ & $0.44\pm 0.19$ & $2.14$ & $0.75/90$ \\
Aug 10    & $2.58\pm 0.07$ & $0.24\pm 0.22$ & $2.08$ & $0.87/69$ \\
Aug 12    & $2.8\pm 0.2$   & $<0.63$        & $1.27$ & $1.18/16$ \\
Aug 20    & $2.4\pm 0.2$   & $<0.65$        & $1.14$ & $1.14/24$ \\
Aug 22    & $2.8\pm 0.2$   & $0.89\pm 0.62$ & $1.42$ & $1.20/24$ \\
\hline
\end{tabular}
\caption{Summary for fits with \emph{Swift}/XRT data}
\label{tab:a}
\end{table}

\begin{figure}
  \includegraphics[angle=270,scale=0.35]{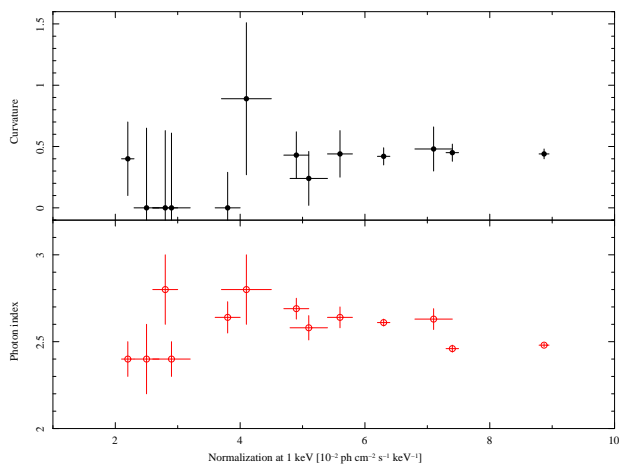}
  \caption{Curvature ($b$) and photon index (or spectral slope $a$) versus model normalization ($K$) of the log-parabolic model of Eq. (1) applied to \emph{Swift}/XRT data in the $0.3-10$ keV energy range.}
\end{figure}

In Fig. 1 we display the curvature (\emph{top panel}) and the photon index (\emph{bottom panel}) plotted versus the normalization. The constant trend revealed by the data confirms the findings reported in \cite{Foschini:2007}: there was no significant spectral changes in the X-ray emission from PKS $2155-304$ during the outburst (corresponding to increasing X-ray fluxes in Fig. 1).

\bibliographystyle{aipprocl} 

\end{document}